\documentclass[prd,aps]{revtex4}

\usepackage{graphicx,subfigure}

\newcommand{\be}{\begin{equation}}
\newcommand{\ee}{\end{equation}}
\newcommand{\bea}{\begin{eqnarray}}
\newcommand{\eea}{\end{eqnarray}}
\newcommand{\bd}{\begin{displaymath}}
\newcommand{\ed}{\end{displaymath}}

\newcommand{\ad }{a^{\dagger}}

\newcommand{\ai}{a_i^{\dagger}}
\newcommand{\aj}{a_j^{\dagger}}
\newcommand{\ab}{a_1^{\dagger}}
\newcommand{\at}{a_2^{\dagger}}

\newcommand{\n }{ | n_1, n_2  >}
\newcommand{\np }{ | n_1 +1 , n_2  >}
\newcommand{\nm }{ | n_1 - 1  , n_2  >}
\newcommand{\nnp }{ | n_1   , n_2 +1  >}
\newcommand{\nnm }{ | n_1   , n_2 -1  >}

\newcommand{\na }{ \sum_{k=1}^d n_k }
\newcommand{\nao }{ \sum_{k=1}^d N_k }

\newcommand{\vn }{ | \vec{n} >}
\newcommand{\nim }{ | \vec{n} - n_i \vec{e}_i >}
\newcommand{\nip }{ | \vec{n} + n_i \vec{e}_i >}

\newcommand{\nul }{ | \vec{0} >}

\newcommand{\hh}{ \hat{H} }

\newcommand{\kk}{ \kappa }
\newcommand{\ak}{ A_{ \{ \kappa \} } (1) }
\newcommand{\akt}{ A_{ \kappa  } (2) }

\begin{document}

%\draft
%\preprint{
%\begin{tabular}{l}
%\hbox to\hsize{\hfill KAIST-TH 2006/10}\\
%[-1mm]
%\hbox to\hsize{\hfill KIAS-P06047}\\
%[-2mm] \hbox to\hsize{\hfill hep-ph/yymmdd}\\
%[-3mm] \hbox to\hsize{\hfill November 2011}\\
%[-3mm]
%\end{tabular}
%}

\title{
On the representation of $A_{\kk}(2)$ algebra and $A_{\kk}(d)$ algebra
}

\author{ Won Sang Chung }
\email{mimip4444@hanmail.net}

\affiliation{
Department of Physics and Research Institute of Natural Science, College of Natural Science, Gyeongsang National University, Jinju 660-701, Korea
}

\date{\today}

\begin{abstract}

In this paper the representation of $A_{\kk}(2)$ algebra given by Daoud and Kibler [M.Daoud and M.Kibler , J.Phys.A{\bf 43} 115303 (2010) , {\bf 45} 244036 (2012) ] is investigated. It is shown that the new generators are necessary for consistency of the algebra. The multi-mode extension of $A_{\kk}(2)$ algebra , which is called $A_{\kk}(d)$ algebra, is also obtained. The positivity condition of the energy spectrum is also investigated for $A_{\kk}(2)$ algebra and $A_{\kk}(d)$ algebra.

\end{abstract}

\maketitle

\section{Introduction}

In two decades, many deformations of a boson algebra have been accomplished. Some of them are constructed by using the Jackson's $q$- calculus [1], while others are not. The deformed boson algebra through $q$- calculus is called a q-boson algebra, which was firstly accomplished by Arik and Coon [2] and lately accomplished by Macfarlane [3] and Biedenharn [4] by using the q-calculus which was originally introduced by Jackson in the early 20th century [1]. In the study of the basic hypergeometric function Jackson invented the Jackson derivative and integral, which is now called q-derivative and q-integral. Jackson's pioneering research enabled theoretical physicists and mathematician to study the new physics or mathematics related to the q-calculus. Much was accomplished in this direction and work is under way to find the meaning of the deformed theory.

Recently Daoud and Kibler [5,6,7] introduced the most interesting algebra which is not related to the $q$-calculus. Their algebra is called a $ \ak$ algebra or a generalized Weyl-Heisenberg algebra. It is a polynomial algebra which depends on some parameters. The $\ak$ algebra is defined as
\be
[a, \ad ] = N \prod_{i=1}^r [ 1 + \kappa_i ( N-1) ], ~~~ [ N, a] = -a , ~~~ [N, \ad ] = \ad, 
\ee
where $\{ k \}$ implies the set of r real parameters as follows :
\be
\{ k \} = \{ \kk_1, \kk_2 , \cdots, \kk_r \}
\ee
The representation of this algebra is well known in the ref [5,6,7]. They obtained a finite dimensional representation and an infinite dimensional representation for this algebra. They also considered the extension of $ A_{\kk}(1) $ (with one degree of freedom ) to $A_{\kk}(2)$ involving two degrees of freedom. They insisted that the  $A_{\kk}(2)$ algebra is spanned by $ a, \ad $ and $N_i $ satisfying
\bd
[ a_i , \ai ] = 1 + \kk ( N_1 + N_2 + N_i ) , ~~~ [N_i , a_j ] = - \delta_{ij} a_j , ~~~  [N_i , \aj ] = - \delta_{ij} \aj ,
\ed
\be
[a_i , a_j ] =0, ~~~ [\ai , \aj ] =0 , ~~~ [a_i , [a_i , \aj ]] = [ \ai , [\ai , a_j ] ]=0, ~~~( i \ne j ),
\ee
where $\kk$ is real and $i, j  \in \{ 1, 2 \} $.
However, Daoud and Kibler did not find the concrete form of the commutation relation between $ a_i $ and $ \aj $ for $ i \ne j $ case.

In this paper we will construct the concrete form of the commutation relation between $ a_i $ and $ \aj $ for the  $A_{\kk}(2)$ algebra. In this process, it is shown that the new generators , which I call $X_{12}$ and $ X_{21} $,  are necessary. Moreover, we will extend our work to the $A_{\kk}(d)$ algebra including $d$ degrees of freedom.

\section{ Representation of $\akt$ algebra }

In this section, we will find the representation of $\akt$ algebra. To do so, let us introduce the eigenvector of two number operators as follows :
\be
N_1 \n = (n_1 +\nu) \n , ~~~ N_2 \n = (n_2 +\nu ) \n , ~~~ n_1, n_2 = 0, 1, 2, \cdots,
\ee
where $N_1, N_2 $ are hermitian and the ground state $|0,0>$ obeying
\be
a_1 |0,0> = a_2 |0,0> = 0
\ee
From the algebra (3), we have following representation
\bd
a_1 \n = f_1 ( n_1 , n_2 ) \nm, ~~~ \ab \n = f_1 ( n_1 +1 , n_2 ) \np
\ed
\be
a_2 \n = f_2 ( n_1 , n_2 ) \nnm, ~~~ \at \n = f_2 ( n_1 , n_2 +1  ) \nnp
\ee
Inserting the eq.(6) into the first relation of the eq.(3), we have
\bd
f_1^2 ( n_1 +1 , n_2 ) - f_1^2 ( n_1  , n_2 ) = 1 + \kk ( 2(n_1 + \nu  ) + n_2 + \nu ) ,
\ed
\be
f_2^2 ( n_1  , n_2 +1 ) - f_2^2 ( n_1  , n_2 ) = 1 + \kk ( n_1 + \nu   + 2( n_2 + \nu )  ) ,
\ee
Solving the recurrence relation (7), we have
\bd
f_1^2 ( n_1  , n_2 ) = \kk (n_1 + \nu )^2 + \kk (n_1 + \nu) (  n_2 + \nu ) + ( 1- \kk ) (n_1 + \nu ) + g_1 ( n_2 ) ,
\ed
\be
f_2^2 ( n_1  , n_2 ) = \kk (n_2 + \nu )^2 + \kk (n_1 + \nu) (  n_2 + \nu ) + ( 1- \kk ) (n_2 + \nu ) + g_2 ( n_1 ) ,
\ee
where $g_1 ( n_2 ) ,  g_2 ( n_1 ) $ are arbitrary functions.

To factorize the right hand sides of the eq.(8), we choose
\be
g_1 ( n_2) = (1-\kk ) n_2 , ~~~ g_2 ( n_1 ) = (1-\kk ) n_1
\ee
Then we have the following representation :
\bd
a_1 \n = \sqrt{ ( n_1 + n_2 + 2 \nu )( \kk ( n_1 + \nu ) + 1 - \kk ) }  \nm,
\ed
\bd
\ab \n = \sqrt{ ( n_1 + n_2 + 2 \nu +1 )( \kk ( n_1 + \nu ) + 1 ) } \np
\ed
\bd
a_2 \n = \sqrt{ ( n_1 + n_2 + 2 \nu )( \kk ( n_2 + \nu ) + 1 - \kk ) } \nnm,
\ed
\be
\at \n = \sqrt{ ( n_1 + n_2 + 2 \nu +1 )( \kk ( n_2 + \nu ) + 1  } \nnp
\ee
and the operator relation is given by
\be
\ai a_i = (N_1 + N_2 ) ( \kk N_i + 1 - \kk )
\ee
Because $|0,0>$ is annihilated by $a_1, a_2 $, we have
\bd
\nu = 1 - \frac {1}{\kk }
\ed
For the representation (10), we can easily check that
\be
[a_1 , a_2 ] =0, ~~~ [\ab , \at ] =0
\ee

From the fact that
\be
[a_2 , \ab a_1 ] = ( \kk N_1 + 1 - \kk ) a_2 , ~~~ [a_1 , \at a_2 ] = ( \kk N_2 + 1 - \kk ) a_1 , 
\ee
we should impose the following relation :
\be
[a_2 , \ab ] = X_{21},~~~[a_1 , \at ] = X_{12} , 
\ee
where the new operator $X_{21}$ and $X_{12}$ are defined as
\be
X_{21} a_1 = ( \kk N_1 + 1 - \kk ) a_2, ~~~X_{12} a_2 = ( \kk N_2 + 1 - \kk ) a_1
\ee
and $X_{21} = X_{12}^{\dagger} $. The new generators $X_{21}$ and $X_{12}$ also satisfy the following commutation relation:
\bd
\ab X_{12}  = \at ( \kk N_1 + 1 - \kk ) , ~~~\at X_{21}  = \ab ( \kk N_2 + 1 - \kk ) ,
\ed
\be
[X_{12}, a_1 ] = [ X_{21}, \ab ] = [X_{21}, a_2 ] = [X_{12}, \at  ] =0
\ee

Acting the eq.(15) on the eigenvector of the number operators, we obtain the matrix representation of $X_{21}$ and $X_{12}$ as follows :
\bd
X_{21} \n = \sqrt{ ( \kk ( n_1 + \nu ) +1 ) ( \kk ( n_2 + \nu ) + 1 -\kk  } |n_1+1, n_2-1>,
\ed
\be
X_{12} \n = \sqrt{ ( \kk ( n_1 + \nu ) +1 -\kk ) ( \kk ( n_2 + \nu ) + 1  } |n_1-1, n_2+1>,
\ee
It can be easily checked that the relation (14) obey the following :
\be
[a_i , [a_i , \aj ]] = [ \ai , [\ai , a_j ] ]=0, ~~~( i \ne j )
\ee

Now let us consider the deformed harmonic potential problem. If we define the deformed harmonic Hamiltonian $H $ as
\be
H = X_1^2 + X_2^2 + P_1^2 + P_2^2 , 
\ee
we have
\be
H  = \kk ( N_1 + N_2)^2 + \left( 2 - \frac{\kk}{2} \right)(N_1 + N_2 ) +1
\ee
Acting the hamiltonian on the number eigenvectors , we have
\be
H |n_1, n_2>  = E_{n_1, n_2}|n_1, n_2> 
\ee
where 
\be
E_{n_1, n_2} =  \kk ( n_1 + n_2  )^2 + ( 2- \frac{\kk }{2} ) (n_1 + n_2 ) + 1 
\ee
The energy  eigenvalue $E_{n_1, n_2}$ depends on the value of $ n_1 + n_2 $. The ground state $|0,0>$ is non-degenerate, but all the excited states are degenerate and the degeneracy is $ {}_2H_n = {}_{n+1}C_n $. If we set $ n_1 + n_2 = n $, the energy eigenvalue is defined as
\be
E_{n} =  \kk n^2 + ( 2- \frac{\kk }{2} ) n + 1,
\ee
where $ n \ge 0 $. 

Now let us discuss the positivity condition of the energy eigenvalue. Because the energy of the harmonic Hamiltonian is positive in an ordinary quantum mechanics. Thus we should demand the positivity of the energy spectrum for the deformed harmonic Hamiltonian. This problem depends on the sign of $\kk$.

\vspace{1cm}

{\bf Case I : $ \kk < 0 $}

\vspace{1cm}

In this case, the positivity of the energy is not guaranteed for all $n$. 

\vspace{1cm}

{\bf Case II : $ \kk = 0 $}

\vspace{1cm}

In this case, the positivity of the energy is guaranteed for all $n$. 

\vspace{1cm}

{\bf Case III : $ \kk > 0 $}

\vspace{1cm}

In this case , the positivity of the energy is guaranteed for all $n$ only when $ \kk \ge 12- 8 \sqrt2 $ .

\section{Representation of $A_{\kk}(d)$ algebra}

In this section we will extend the result of section II to the multi-mode case. We will set the number of modes to $d$. Then the algebra has $d$ degrees of freedom and we will denote this algebra by  $A_{\kk}(d)$ algebra. The  $A_{\kk}(d)$ algebra is defined as
\bd
[ a_i , \ai ] = 1 + \kk ( \nao + N_i ) , ~~~ [N_i , a_j ] = - \delta_{ij} a_j , ~~~  [N_i , \aj ] = - \delta_{ij} \aj ,
\ed
\be
[a_i , a_j ] =0, ~~~ [\ai , \aj ] =0 , ~~~ [a_i , [a_i , \aj ]] = [ \ai , [\ai , a_j ] ]=0, ~~~( i \ne j ),
\ee
where $\kk$ is real and $i, j  \in \{ 1, 2, \cdots , d  \} $.

Let us introduce the eigenvector of number operators as follows :
\be
N_i \vn = (n_i +\nu) \vn ,~~~ n_i = 0, 1, 2, \cdots,
\ee
where $N_i $ are hermitian and the ground state $\nul$ obeying
\be
a_i \nul = 0
\ee
and $\vn$ denotes $|n_1, n_2, \cdots, n_d >$ and $\nul$ denotes $|0,0,\cdots, 0 > $.

From the algebra (24), we have following representation
\bd
a_i \vn = \sqrt{ ( \na  + d \nu )( \kk ( n_i + \nu ) + 1 - \kk ) }  \nim,
\ed
\be
\ai \vn = \sqrt{ ( \na  + d \nu +1  )( \kk ( n_i + \nu ) + 1 ) }  \nip,
\ee
where $\nim$ and $\nip$ is defined as follows :
\bd
\nim =|n_1, n_2, \cdots , n_i -1 , n_{i+1} , \cdots , n_d >
\ed
\be
\nip =|n_1, n_2, \cdots , n_i +1 , n_{i+1} , \cdots , n_d >
\ee
and the operator relation is given by
\be
\ai a_i  = (\nao ) ( \kk N_i + 1 - \kk )
\ee
Because $\nul $ is annihilated by $a_i $'s, we have
\bd
\nu = 1 - \frac {1}{\kk }
\ed

For the representation (27), we can easily check that
\be
[a_i , a_j ] =0, ~~~ [\ai , \aj ] =0
\ee

From the fact that
\be
[a_j , \ai a_i ] = ( \kk N_i + 1 - \kk ) a_j , 
\ee
we should impose the following relation :
\be
[a_j , \ai ] = X_{ji},
\ee
where the new operator $X_{ji}$ is defined as
\be
X_{ji} a_i = ( \kk N_i + 1 - \kk ) a_j 
\ee
and $X_{ji}^{\dagger} = X_{ij} $.
The new generators $X_{ji}$ also satisfy the following commutation relation:
\be
\ai X_{ij}  = \aj ( \kk N_i + 1 - \kk ) , ~~~ [X_{ji}, a_j ] = [ X_{ij}, \aj ] =0
\ee
Acting the eq.(34) on the eigenvector of the number operators, we obtain the matrix representation of $X_{ji}$ as follows :
\be
X_{ji} \vn = \sqrt{ ( \kk ( n_i + \nu ) +1 ) ( \kk ( n_j + \nu ) + 1 -\kk  } | \vec{n} + \vec{e}_i - \vec{e}_j >
\ee
It can be easily checked that the relation (32) obey the following :
\be
[a_i , [a_i , \aj ]] = [ \ai , [\ai , a_j ] ]=0, ~~~( i \ne j )
\ee

If we define the deformed harmonic Hamiltonian $\hh $ as
\be
H = \sum_{i=1}^d ( X_i^2 + P_i^2 ), 
\ee
we have
\be
H  = \kk ( \nao )^2 + \left( d- \frac{d-1}{2}k \right) \nao + \frac {d}{2}
\ee
Acting the hamiltonian on the number eigenvectors , we have
\be
H \vn = E_{\vec{n}}\vn , 
\ee
where
\be
E_{\vec{n}}= \left( \kk ( \na )^2 + ( d- \frac{d-1}{2}k ) \na + \frac {d}{2} \right) \vn
\ee
The energy  eigenvalue $E_{\vec{n}}$ depends on the value of $ \na  $. The ground state $\nul$ is non-degenerate, but all the excited states are degenerate and the degeneracy is $ {}_d H_n = {}_{d+n-1}C_n $. If we set $ \na  = n $, the energy eigenvalue is defined as
\be
E_{n} =  \kk n^2 + \left( d- \frac{d-1}{2}k \right)  n + 1,
\ee
where $ n \ge 0 $.

Now let us discuss the positivity condition of the energy eigenvalue. Because the energy of the harmonic Hamiltonian is positive in an ordinary quantum mechanics. Thus we should demand the positivity of the energy spectrum for the deformed harmonic Hamiltonian. This problem depends on the sign of $\kk$.

\vspace{1cm}

{\bf Case I : $ \kk < 0 $}

\vspace{1cm}

In this case, the positivity of the energy is not guaranteed for all $n$.

\vspace{1cm}

{\bf Case II : $ \kk = 0 $}

\vspace{1cm}

In this case, the positivity of the energy is guaranteed for all $n$.

\vspace{1cm}

{\bf Case III : $ \kk > 0 $}

\vspace{1cm}

In this case , the positivity of the energy is guaranteed for all $n$ only when $ \kk \ge \frac{2d(d+1) - 4d \sqrt{d} }{(d-1)^2 } $ .

\section*{Conclusion}

In this paper we discussed the representation of $A_{\kk}(2)$ algebra given by Daoud and Kibler [5, 6, 7]. We found that the new generators are necessary for consistency of the algebra. We extended $A_{\kk}(2)$ algebra to the multi-mode case , which is called $A_{\kk}(d)$ algebra. We also found the positivity condition of the energy spectrum for the deformed harmonic Hamiltonian for $A_{\kk}(2)$ algebra and $A_{\kk}(d)$ algebra.

%%%%%%%%%%%%%%%%%% References
%%%%%%%%%%%%%%%%%%%%%%%%%%%%%%%%%%%%%%%%%%%%%%%%%%%%%%%%%%%%%%%%%%%%%%%
\def\JMP #1 #2 #3 {J. Math. Phys {\bf#1},\ #2 (#3)}
\def\JP #1 #2 #3 {J. Phys. A {\bf#1},\ #2 (#3)}
\def\JPD #1 #2 #3 {J. Phys. D {\bf#1},\ #2 ( #3)}
\def\PRL #1 #2 #3 { Phys. Rev. Lett. {\bf#1},\ #2 ( #3)}

%%%%%%%%%%%%%%%%%%%%%%%%%%%%%%%%%%%%%%%%%%%%%%%%%%%%%%%%%%%%%%%%%%%%%%%

\section*{Refernces}
[1] F. Jackson, Mess.Math. {\bf 38 }, 57 (1909).

[2] M.Arik and D.Coon, \JMP 17 524 1976 .

[3] A.J.Macfarlane, \JP 22 4581 1989 .

[4] L.Biedenharn, \JP 22 L873 1990 .

[5] M.Daoud and M.Kibler , \JP 43 115303 2010 .

[6] M.Daoud and M.Kibler, \JP 45 244036 2012 .

[7] M.Daoud and M.Kibler , \JMP 52 081201 2011 .

\end{document}